\documentclass[twocolumn,floatfix,preprintnumbers]{revtex4-1}

\usepackage[utf8]{inputenc}
\usepackage[colorlinks=true,citecolor=blue,linkcolor=blue]{hyperref}
\usepackage[normalem]{ulem}
\usepackage{url}
\usepackage{graphicx,wrapfig,float,slashed,cancel}
\usepackage{amsmath,amssymb,epsfig,graphicx,xcolor,stmaryrd}
\usepackage{bm}
\usepackage{enumitem}

\definecolor{darkblue}{RGB}{1, 90, 173}

\begin{document}


\title{The newly observed $ Z_{cs}(3985)^- $ state: in vacuum and a dense medium}

\author{K. Azizi}
\email{kazem.azizi@ut.ac.ir}
\affiliation{Department of Physics, University of Tehran, North Karegar Ave. Tehran 14395-547, Iran}
\affiliation{ Department of Physics, Do\v{g}u\c{s} University, Acibadem-Kadik\"{o}y, 34722
Istanbul, Turkey}

\author{N. Er}
\email{nuray@ibu.edu.tr}
\affiliation{ Department of Physics, Bolu  Abant \.{I}zzet Baysal University,
G\"olk\"oy Kamp\"us\"u, 14030 Bolu, Turkey}

\date{\today}

\preprint{}

\begin{abstract}
	We report on some properties of the newly observed charged hidden-charmed open strange  $ Z_{cs}(3985)^- $ state by BESIII Collaboration. Assigning the quantum numbers $ J^{P} = 1^{+}$ and the quark composition $ c \bar c s\bar u $ and considering it as the strange partner of the famous $ Z_c(3900) $ state, we estimate the mass of the $ Z_{cs}(3985)^- $ resonance in vacuum and compare it with the experimental data.  We also investigate its mass, current coupling and  vector-self energy in a medium with finite density. Our result on the mass in vacuum agrees well with the experimental data.  We estimate the mass and current coupling of the b-partner of this state, $ Z_{bs}$, in the vacuum as well.  For its mass we get $ m_{Z_{bs}}= 10732^{+97}_{-46}~\mbox{MeV}$, which may be checked via other nonperturbative approaches as well as future experiments.  We present the dependence of the spectroscopic parameters of $ Z_{cs}(3985)^- $  state on density and observe that these parameters are linearly changed  with increasing in the density. 
\end{abstract}


\maketitle

\section{Introduction} \label{sec:intro}
The recently discovered charged  resonances are good candidates of tetraquark systems as they can not be put in the spectrum of the usual  $ c \bar c  $ mesons by any means.  The observation of the ground state $Z_c(3900)$ was reported  by the BESIII  \cite{Ablikim:2013mio}  and  Belle \cite{Liu:2013dau}  Collaborations in 2013, simultaneously. After these observations many theoretical studies were performed on different properties of these states (as examples see  Refs. \cite{Agaev:2016dev,PhysRevD.96.034026,Agaev:2020zad,Azizi:2020itk,Ozdem:2017jqh}  and references therein).  The next state in this group is the $ Z(4430) $ which discovered by the Belle and LHCb Collaborations \cite{PhysRevLett.100.142001,PhysRevD.88.074026,PhysRevD.90.112009,PhysRevLett.112.222002,PhysRevD.92.112009} and fully studied in Refs. \cite{PhysRevD.96.034026,Lee:2007gs,Branz:2010sh,Wang:2014vha,Ebert:2008kb,Liu:2008qx,Azizi:2020yhs} and references therein. The mass difference between $ Z(4430) $  and $Z_c(3900)$ is roughly equal to the mass difference between $\psi(2S)  $ and $ J/\psi $, that respectively appear in their decay products, was the reason for consideration of $ Z(4430) $ to be the first radial excitation of the $Z_c(3900)$ state \cite{PhysRevD.96.034026,Azizi:2020yhs}. 

Very recently, as the first candidate, the BESIII Collaboration reported the observation of a  charged hidden-charm tetraquark with strangeness, decaying into $D_s^{-}D^{*0}  $ and $ D_s^{*-} D^0 $ in  the process $ e^+e^- \rightarrow K^+ (D_s^{-}D^{*0} +D_s^{*-} D^0 )$ \cite{Ablikim:2020hsk}. This new member is called $ Z_{cs}(3985)^- $ and its  quark composition is most likely   $ c \bar c s\bar u $.  The  measured mass is  just above the thresholds of  $ D_s \bar D^* $ and  $ D^*_s \bar D $.  This value together with the value of the width reported by the experiment are:
\begin{eqnarray}
m&=&3982.5^{+1.8}_{-2.6}\pm 2.1~~ \mbox{MeV},\\ \nonumber
\Gamma&=&12.8^{+5.3}_{-4.4}\pm 3.0~~~~~ \mbox{MeV}.
\end{eqnarray}
The charmonium-like open strange four quark systems were already in agenda of different theoretical studies \cite{Voloshin:2019ilw,Dias:2013qga,Lee:2008uy,Chen:2013wca,Ferretti:2020ewe}. In Ref. \cite{Voloshin:2019ilw}, assuming SU(3) flavor symmetry,  the existence of strange partners to the $ Z_c $ states denoted by $ Z_{cs} $ was predicted. In this study, as partners of $Z_c(4100)$ and $Z_c(4200)$ hadrocharmonium states, the hypothetical strange hadrocharmonium resonances  with quantum numbers $J^P=0^+$ and $1^+$  were referred as $Z_{cs}(4250)$ and $Z_{cs}(4350)$, respectively.  In Ref. \cite{Dias:2013qga}, the mass of the the possible  $D_sD^*$ molecular state, as the strange partner of the observed $Z_c^{\pm}(3900)$ states, called $Z_{cs}^+$ was predicted.   In Ref. \cite{Lee:2008uy}, using the QCD sum rule approach, and considering a state near to the $D_s\bar{D}^*/D_s^*\bar{D}$ threshold, its four different decay channels were studied and by calculating the related strong coupling constants and  decay widths, the total width of the proposed state was predicted. In Ref. \cite{Chen:2013wca}, while studying  the hidden-charm di-kaon decays of higher charmonia and charmonium-like state, the authors predicted the existence of hidden-charm and open-strange channels near the $D\bar{D}_s^*/D^*\bar{D}_s$ and $D^*\bar{D}_s^*/\bar{D}^*D_s^*$ thresholds. In Ref. \cite{Ferretti:2020ewe}, the authors calculated the spectrum of strange hidden-charm and bottom tetraquarks in the hadro-quarkonium and the compact tetraquark models.

 However, there  appeared various studies after the BESIII Collaboration observation report. Different assumptions and assignments have been made to clarify the nature and structure of the observed state \cite{Wang:2020kej,Meng:2020ihj,Liu:2020nge,Yang:2020nrt,Chen:2020yvq,Du:2020vwb,Cao:2020cfx,Sun:2020hjw,Wang:2020rcx}. In Ref. \cite{Wang:2020kej},   by performing a combined analysis for the line shape of the recoil mass distribution of $K^+$ at five energy points, the authors find that the $Z_{cs}(3985)^-$ can be explained as a reflection structure of charmed-strange meson $D^*_{s2}(2573)$, which is produced from the open-charm decay of $Y(4660)  $ with a $ D^*_s $ meson. In Ref. \cite{Meng:2020ihj},  the authors introduced the $G_{U/V}$ parity to construct the flavor wave functions of the $Z_{cs}$ states and predicted their masses. They also predicted the masses of the tetraquark states $Z_{bs}$ and $Z'_{bs}$, which are the $U/V$ -spin partner states of $Z_b(10610)^{\pm}$ and $Z_b(10650)^{\pm}$, respectively. In Ref. \cite{Liu:2020nge}, using the one-boson-exchange model, it was concluded that the $Z_{cs}(3985)$ could not be a pure hadronic molucular state of $\bar{D}_sD^*$ and $\bar{D}^*_sD$ and could consist of large components of compact nature. In Ref. \cite{Yang:2020nrt}, the quantum numbers $I(J^P)=\frac{1}{2}(1^+)$ were assigned to the  new resonance $Z_{cs}$ by the mass analyses and the comparision of the results with the experimental data. After producing the phase shifts for all the considered channels, the results of Ref. \cite{Chen:2020yvq},  exclude the state $Z_{cs}(3985)^-$  as a $D^{*0}D_s^{-}/D^{0}D_s^{-}/D^{*0}D_s^{*-}$ resonance with the quantum numbers $I(J^P)=1/2(1^+, 0^-, 1^-, 2^-)$. In Ref. \cite{Du:2020vwb}, the states $Z_c(3900)  $ and $ Zc(4020) $, as well as their strange partners $  Z_{cs}(3985) $ and $Z_{cs}(4120)  $ were studied. With an overall short-ranged contact potential in SU(3) flavor symmetry  the authors showed that the pole structures of these states could be extracted from the experimental data. In Ref. \cite{Cao:2020cfx}, the authors studied  the photoproduction of exotic states $Z_{cs}(3985)$ and $Z'_{cs}$ and explored the possibility of searching for them in future electron-ion colliders (EIC). In Ref \cite{Sun:2020hjw} the authors studied the structure of $Z^-_{cs}(3985)$ state in the picture of $D^{(*)-}_sD^{(*)0}$ molecular state and got a consistent result with the experiment. In Ref. \cite{Wang:2020rcx}, the authors calculated the mass spectra for the $\bar{D}^{(*)}_sD^{(*)}$ molecular states and $sc\bar{q}\bar{c}$ tetraquark states with $J^P=0^+, 1^+, 2^+)$ and got consistent results with the experiment in both scenarios. In both the molecular and diquark-antidiquark pictures, their results suggested that there may exist two almost degenerate states, as the strange partners of the $ X(3872) $ and $ Zc(3900) $.

In the present study, we investigate the  $ Z_{cs}(3985)^- $ (hereafter $ Z_{cs} $) state   both in the vacuum and a medium with finite density. Based on the provided information by the experiment that the new state is composed of  $ c \bar c s\bar u $ and the measured central value for its mass is sightly higher  than  the thresholds $D_s\bar{D}^{*}$ and   $D^*_s\bar{D}$, the $ Z_{cs} $ state can be considered as a compact tetraquark rather than the loosely bound molecular states $D_s\bar{D}^{*}$ and   $D^*_s\bar{D}$. With this information and  the results of  theoretical studies before the experimental report, Refs. \cite{PhysRevD.96.034026,Azizi:2020itk,Voloshin:2019ilw,Dias:2013qga,Lee:2008uy,Chen:2013wca,Ferretti:2020ewe},    we treat  the $ Z_{cs} $ as a compact tetraquark and  assign the quantum numbers $ J^{P} = 1^{+}$ and quark composition $ c \bar c s\bar u $ to this state  and consider it as the strange partner of the famous $ Z_c(3900) $ state. We calculate its mass and current coupling in vacuum. We also investigate this particle in a dense medium to estimate its vector self-energy, mass and current coupling at nuclear matter saturation density and study the changes of its parameters with respect to the density of the medium in order to more clarify its nature and structure. We use the technique of the QCD sum rules \cite{SHIFMAN1979385,SHIFMAN1979448} to calculate the spectroscopic parameters of this state. Our calculations let us also estimate some parameters of the b-partner of  $ Z_{cs} $ state, namely  $ Z_{bs} $, which may be in agenda of future experiments.

 In in-medium sum rules,  the medium effects are included via various operators in the nonperturbative parts of the operator product expansion (OPE)  (see for example \cite{COHEN1995221,Azizi:2014yea}). The new progresses recorded in the heavy ion collision experiments like  the Relativistic Heavy Ion Collider (RHIC)  at the Bookhaven National Laboratory (BNL) and the Large Hadron Collider (LHC) and  some in-medium experiments like  $ \bar{P}ANDA $ at Facility for Antiproton and Ion Research (FAIR) at GSI and  the Nuclotron-based Ion Collider Facility at JINR (NICA) at Dubna,   it  will be  possible to study not only the  ground state particles but also the excited exotic states  \cite{Cho:2017dcy,Cho:2011ew,Wiedner:2011mf}.   Hence, these experiments  may provide good opportunities to investigate  $Z_c(3900)$,  $Z(4430)$ and $ Z_{cs} $ in order to clarify their nature and substructure and fix their quantum numbers. 

 In next section we provide some details of the calculations to obtain sum rules for the physical quantities under study. Section III encompasses the numerical analyses of the  obtained expressions. The last section is devoted to the summary of the results and conclusions. 


\section{Sum rules for in-medium parameters of $ Z_{cs} $}

The information provided by the BESIII collaboration lead us to assign the quantum numbers $J^{P}=1^{+}$ to $ Z_{cs} $ state and  consider it as the strange partner of the famous  $Z_c (3900)$ particle with quark composition $ c \bar c s\bar u $. The following current can interpolate the  $ Z_{cs} $ state with the assumed nature: 
\begin{eqnarray}\label{ }
J_{\mu}(x)&=&\frac{i\epsilon_{abc}\epsilon_{dec}}{\sqrt{2}} \Bigg \{ \Big[s_{a}^T(x) C \gamma_5 c_b(x)\Big] \Big[\bar{u}_{d}(x) \gamma_{\mu} C\bar{c}^T_e(x) \Big] \nonumber \\
&-& \Big[s_{a}^T(x) C \gamma_{\mu} c_b(x)\Big] \Big[\bar{u}_{d}(x) \gamma_5 C\bar{c}^T_e(x) \Big] \Bigg \},
\end{eqnarray}
where $\epsilon_{abc}$ and $\epsilon_{dec}$ are anti-symmetric Levi-Civita symbols with  $a, b, c, d$ and $e$ being color indices. Here,  $T$  stands for the  transpose and $C$  represents the charge conjugation operator.

The basic function that we need to evaluate is the  following  in-medium correlation function (CF): 
\begin{equation}\label{corre1}
\Pi_{\mu\nu}(p)=i\int{d^4 xe^{ip\cdot x}\langle\psi_0|\mathcal{T}[J_{\mu}(x)J^{\dagger}_{\nu}(0)]|\psi_0\rangle},
\end{equation}
where $p$ is the four-momentum of the $Z_{cs}$ state, $\mathcal{T}$ is the time ordering operator,  and $|\psi_0\rangle$ is the  ground state of  nuclear medium.  We will calculate this two-point CF once in terms of in-medium hadronic parameters and the second in terms of QCD fundamental parameters and the in-medium quark and gluon condensates. By matching the obtained results in two different pictures we will able to calculate the in-medium hadronic parameters in terms of the QCD degrees of freedom.  We should note that although the QCD sum rule method \cite{SHIFMAN1979385,SHIFMAN1979448} was originally formulated for calculations of different parameters of the standard mesons and baryons, it has been successfully applied for the tetraquarks and other multiquark systems giving consistent results with the experiments on the observed states.  This approach  has helped physicists to describe the results of experiments on the XYZ tetraquark systems provided by LHCb, BESIII, Belle and other related experiments and fix the nature, quark-gluon organization and quantum numbers of these states (for some of these studies see for instance \cite{Agaev:2016dev,PhysRevD.96.034026,Agaev:2020zad,Azizi:2020itk,Chen:2018kuu,Matheus:2006xi,Chen:2015fsa,Sundu:2018nxt,Wang:2020xyc,Wang:2018qpe} and references therein). Despite the successes of the QCD sum rules in applications to the tetraquark systems, there are some doubts on the applied approaches to the tetraquark molecular states. In Refs. \cite{Lucha:2019pmp,Lucha:2019cpe}, it is stated that the contributions at the orders ${\cal O} (1) $ and ${\cal O} (\alpha_s) $ in OPE are exactly canceled out by the meson-meson scattering states at the hadronic side and the tetraquark molecular states begin to receive contributions at the order ${\cal O }(\alpha^2_s)$. However, in  Ref. \cite{Wang:2020cme}, by choosing an axialvector  and a tensor currents, it is shown that  the meson-meson scattering states do not give contribution to the CF, while the tetraquark molecular states  can saturate it. It is also stated that the Landau equation is of no use to study the Feynman diagrams in the QCD sum rules for the tetraquark molecular states and the OPE of these states  receive contributions at leading orders.

The hadronic representation of CF is  obtained by saturating it with a  full set of hadronic state. By separating the ground state contribution, we get the following expression after performing the four-intergral over coordinate space:
\begin{equation} \label{had1}
\Pi^{Had}_{\mu\nu}(p)=- \frac{\langle\psi_0|J_{\mu}|Z_{cs}(p)\rangle \langle Z_{cs}(p)|J^{\dagger}_{\nu}|\psi_0\rangle}{p^{*2}-m_{Z_{cs}}^{*2}} + ... ,
\end{equation}
where $p^{*}$ is the in-medium momentum, $ m_{Z_{cs}}^{*} $ is the in-medium mass and  dots stand for the  contributions of the  higher  states and  continuum. The above expression can be further simplified by introducing the in-medium decay constant or current coupling   $  f^{*}_{Z_{cs}} $  defined by 
\begin{equation} \label{had2}
\langle\psi_0|J_{\mu} |Z_{cs}(p)\rangle = f^{*}_{Z_{cs}} m_{Z_{cs}}^{*} \varepsilon_{\mu},
\end{equation}
where $\varepsilon_{\mu}$ is  the polarization four-vector  of $Z_{cs}$.  Using this definition leads to the summation over the polarization vector. Hence, 
\begin{equation}
\Pi_{\mu\nu}^{Had}(p)=-\frac{m^{*2}_{Z_{cs}} f^{*2}_{Z_{cs}}}{p^{*2}-m^{*2}_{Z_{cs}}} \Big[-g_{\mu\nu} +\frac{p_{\mu}^{*} p^{*}_{\nu} }{m^{*2}_{Z_{cs}}} \Big] + ... ,
\end{equation}
is obtained for the hadronic side. 
To proceed, we need to define two kinds of self-energies. First one is  the scalar self-energy defined by the shift in the hadronic mass due to the nuclear medium,  $\Sigma_s=m^{*}_{Z_{cs}}-m_{Z_{cs}}$, which is a Lorentz scalar and   the second one is the  vector self-energy $\Sigma_{\upsilon}$  that transforms as a Lorentz four-vector and denotes the shift in the four-momentum of the particle due to the dense medium. The vector self energy enters to the calculations as $p_{\mu}^*=p_{\mu}-\Sigma_{\upsilon}u_{\mu}$ (for more details about the scalar and vector self energies see \cite{COHEN1995221}).
 Here, $  u_{\mu}$ is the velocity four-vector of the nuclear matter. We  perform the calculations in the rest frame of the medium,  where $u_{\mu}=(1,0)$.  We should remark that  we also choose to work at zero momentum limit of the particle's three momentum,   for which the longitudinal and transverse polarizations become degenerate, and there is no way to distinguish these two polarizations from each other.  For  the  finite and non-zero three momentum in a dense medium, the longitudinal and transverse polarizations are distinguishable and they should be calculated, separately. In terms of the new quantities, the hadronic part takes the form
\begin{eqnarray} \label{PiPhe}
\Pi_{\mu\nu}^{Had}(p)=&-&\frac{ f^{*2}_{Z_{cs}}}{p^2-\mu^2} \Big[-g_{\mu\nu} m^{*2}_{Z_{cs}} +p_{\mu}p_{\nu} \nonumber \\
&-& \Sigma_{\upsilon}p_{\mu}u_{\nu} -\Sigma_{\upsilon}p_{\nu}u_{\mu}+\Sigma_{\upsilon}^2 u_{\mu}u_{\nu} \Big] + ...,\nonumber\\
\end{eqnarray}
where $\mu^2=m^{*2}_{Z_{cs}}-\Sigma^2_{\upsilon} +2p_0\Sigma_{\upsilon}$, with $ p_0=p.u $ being the energy of the quasi-particle in the dense medium.  We apply the  Borel  transformation to suppress the contributions of the higher states and continuum based on the standard prescriptions of the method. As a result, we get
\begin{eqnarray} \label{PiPhe2}
\Pi_{\mu\nu}^{Phe}(M^2)&=&f^{*2}_{Z_{cs}}e^{-\mu^2/M^2} \Big[-g_{\mu\nu} m^{*2}_{Z_{cs}} \nonumber \\
&+& p_{\mu}p_{\nu} - \Sigma_{\upsilon}p_{\mu}u_{\nu} -\Sigma_{\upsilon}p_{\nu}u_{\mu}+\Sigma_{\upsilon}^2 u_{\mu}u_{\nu} \Big]  \nonumber \\&+& ...,
\end{eqnarray}
where $M^2$ is the Borel mass parameter.

The QCD side of the CF is calculated in deep Euclidean region by the help of OPE. In this representation, the short-distance (perturbative) effects are separated from the long-distance (nonperturbative) effects. To this end, we  insert the explicit expression of the interpolating current $J_{\mu}(x)$ into the CF and perform all the possible contractions among the quark fields using the Wick's theorem. As a result, we get an expression in terms of the in-medium light and heavy quark propagators:
\begin{eqnarray}\label{qcd}
&&\Pi_{\mu\nu}^{QCD}(p)= -\frac{i}{2}\varepsilon_{abc} \varepsilon_{a'b'c'} \varepsilon_{dec} \varepsilon_{d'e'c'} \nonumber \\
&\times& \int d^4 x e^{ipx}\Big\{Tr\Big[\gamma_5 \tilde{S}_s^{aa'} (x)\gamma_5 S_c^{bb'}(x)\Big] \nonumber\\
&\times& Tr\Big[\gamma_{\mu} \tilde{S}_c^{e'e} (-x)\gamma_{\nu} S_u^{d'd}(-x)\Big] - Tr\Big[[\gamma_{\mu} \tilde{S}_c^{e'e} (-x) \nonumber\\
&\times& \gamma_5 S_u^{d'd}(-x)\Big]Tr\Big[\gamma_{\nu} \tilde{S}_s^{aa'} (x)\gamma_5 S_c^{bb'}(x)\Big] \nonumber \\
&-& Tr\Big[\gamma_5 \tilde{S}_s^{aa'} (x)\gamma_{\mu} S_c^{bb'}(x)\Big] Tr\Big[\gamma_5 \tilde{S}_c^{e'e} (-x)\gamma_{\nu} S_u^{d'd}(-x)\Big] \nonumber \\
&+& Tr\Big[\gamma_{\nu}  \tilde{S}_s^{aa'} (x)\gamma_{\mu} S_c^{bb'}(x)\Big] \nonumber \\
&\times&Tr\Big[\gamma_5 \tilde{S}_c^{e'e} (-x)\gamma_5 S_u^{d'd}(-x)\Big]  \Big\}_{|\psi_0\rangle},\nonumber \\
\end{eqnarray}
where $\tilde{S}_{q(c)}= C S_{q(c)}^{T}C$, and  $ q $ stands for $ u $,  or $ s $ quark. We make use of the  in-medium light and heavy quark propagators in coordinate space. Against our previous studies on the properties of $ Z_c(3900) $ and $ Z(4430) $ states \cite{Azizi:2020itk,Azizi:2020yhs}, we keep the light quark mass in the calculations in order to take into account especially the strange quark mass effects:
\begin{eqnarray}\label{ne1}
S_q^{ij}(x)&=&
\frac{i}{2\pi^2}\delta^{ij}\frac{1}{(x^2)^2}\not\!x-\frac{m_q}{4\pi^2}\delta^{ij}\frac{1}{x^2}
 + \chi^i_q(x)\bar{\chi}^j_q(0) \nonumber \\
&-&\frac{ig_s}{32\pi^2}F_{\mu\nu}^{ij}(0)\frac{1}{x^2}[\not\!x\sigma^{\mu\nu}+\sigma^{\mu\nu}\not\!x] +\cdots \, ,\\
\mbox{and}\nonumber\\
S_c^{ij}(x)&=&\frac{i}{(2\pi)^4}\int d^4k e^{-ik \cdot x} \left\{\frac{\delta_{ij}}{\!\not\!{k}-m_c}\right.\nonumber\\
&&\left.-\frac{g_sF_{\mu\nu}^{ij}(0)}{4}\frac{\sigma_{\mu\nu}(\!\not\!{k}+m_c)+(\!\not\!{k}+m_c)
\sigma_{\mu\nu}}{(k^2-m_c^2)^2}\right.\nonumber\\
&&\left.+\frac{\pi^2}{3} \Big\langle \frac{\alpha_sGG}{\pi}\Big\rangle\delta_{ij}m_c \frac{k^2+m_c\!\not\!{k}}{(k^2-m_c^2)^4}+\cdots\right\} \, , \nonumber \\
\end{eqnarray}
where $\chi^i_q$ and $\bar{\chi}^j_q$ are the Grassmann background quark fields and we use the short-hand notation
\begin{equation}
\label{ }
F_{\mu\nu}^{ij}=F_{\mu\nu}^{A}t^{ij,A}, ~~~~~A=1,2, ...,8,
\end{equation}
with $F_{\mu\nu}^A$ being the  classical background gluon fields. Here,  $t^{ij,A}=\frac{\lambda ^{ij,A}}{2}$ and  $
\lambda ^{ij, A}$ are the  Gell-Mann matrices. 

Now, we insert the explicit forms of the light and heavy quark propagators into the CF (\ref{qcd}) and perform all the four-integrals over $ x  $ to go to the momentum space. We then perform the Borel transformation in Euclidean region and apply the continuum subtraction procedure. Here, we do not present the details of calculations, for which one may refer to Refs. \cite{Azizi:2020itk,Azizi:2014yea,Azizi:2018duk}. After these standard but lengthy procedures, one gets the QCD side of the CF in terms of different Lorentz structures as
\begin{eqnarray}\label{QCDcof}
&&\Pi_{\mu\nu}^{QCD}(M^2, s_0^*,\rho)\nonumber \\&=&-\Upsilon^{QCD}_1 (M^2, s_0^*,\rho) g_{\mu\nu}  + \Upsilon^{QCD}_2 (M^2, s_0^*,\rho) p_{\mu}p_{\nu}    \nonumber \\
&-&  \Upsilon^{QCD}_3 (M^2, s_0^*,\rho) p_{\mu}u_{\nu}  -\Upsilon^{QCD}_4 (M^2, s_0^*,\rho) p_{\nu}u_{\mu}  \nonumber \\
&+& \Upsilon^{QCD}_5 (M^2, s_0^*,\rho) u_{\mu}u_{\nu},
\end{eqnarray}
where the  functions $\Upsilon^{QCD}_i$ ($i=1,...,5$) in Eq. (\ref{QCDcof}) are given by 
\begin{equation}\label{Upsil2}
\mathbf{ \Upsilon}^{QCD}_{i}(M^2, s_0^*,\rho)= \int_{(2m_c+m_s+m_u)^2}^{s^{*}_0} ds \psi^{QCD}_{i}(s,\rho)e^{-\frac{s}{M^2}},
\end{equation}
with $s_0^*$ being the in-medium continuum threshold.  The spectral densities $ \psi^{QCD}_{i}(s,\rho) $ are very lengthy functions having various components:  perturbative ($ pert $) as well as  two-quark ($qq$), two-gluon ($gg$) and mixed quark-gluon ($qgq$) condensates, i.e,
\begin{eqnarray}\label{ }
\psi_i^{QCD} (s,\rho)&=&\psi_i^{pert}(s,\rho) + \psi_i^{qq} (s,\rho) + \psi_i^{gg} (s,\rho) \nonumber\\
&+& \psi_i^{qgq} (s,\rho).\nonumber\\
\end{eqnarray}
The explicit forms of the above spectral densities, as an example, corresponding to the  structure $g_{\mu\nu}$ are presented in the
Appendix.

Having completed the hadronic and QCD sides of the calculations, we match the coefficients of different Lorentz structures from the two representations of the CF to construct the sum rules for the physical quantities. They  are obtained as 
\begin{eqnarray}\label{SumR}
-m^{*2}_{Z_{cs}}f^{*2}_{Z_{cs}} e^{-\frac{\mu^2}{M^2}}& = & \mathbf{ \Upsilon}^{QCD}_{1}(M^2, s_0^*,\rho), \nonumber  \\
f^{*2}_{Z_{cs}} e^{-\frac{\mu^2}{M^2}}& = & \mathbf{ \Upsilon}^{QCD}_{2}(M^2, s_0^*,\rho), \nonumber  \\
-\Sigma_{\upsilon}f^{*2}_{Z_{cs}} e^{-\frac{\mu^2}{M^2}}& = &\mathbf{  \Upsilon}^{QCD}_{3}(M^2, s_0^*,\rho), \nonumber  \\
-\Sigma_{\upsilon}f^{*2}_{Z_{cs}} e^{-\frac{\mu^2}{M^2}}& = & \mathbf{ \Upsilon}^{QCD}_{4}(M^2, s_0^*,\rho), \nonumber  \\
\Sigma^2_{\upsilon}f^{*2}_{Z_{cs}} e^{-\frac{\mu^2}{M^2}}& = & \mathbf{ \Upsilon}^{QCD}_{5}(M^2, s_0^*,\rho).
\end{eqnarray}
The expressions obtained for the in-medium mass, current coupling and vector self energy, by this way, depend on the density of the nuclear medium, $ \rho $, in addition to the auxiliary parameters $ M^2 $ and $s_0^*$ and other QCD parameters  \cite{Azizi:2020itk,Azizi:2014yea}. Setting $ \rho\rightarrow 0 $ in first two presented sum rules, we get the expressions for the related physical quantities in vacuum. 

\section{Numerical Results}
The numerical calculations of  the mass, current coupling and vector self-energy  in medium and the mass and current coupling in vacuum  require values of input parameters  such as quark masses,  nuclear matter saturation density,  expectation values of the in-medium quark, gluon and mixed condensates, etc.   They are taken as: $m_u=2.16^{+0.49}_{-0.26}$~MeV, $m_s=93^{+11}_{-5}$~MeV, $m_c=1.27\pm 0.02$~GeV, $\rho^{sat}=0.11^3$~GeV$^3$,  $\langle u^{\dag} u\rangle_{\rho}=\frac{3}{2}\rho$~GeV$^3$, $\langle s^{\dag} s\rangle_{\rho}=0$, $\langle \bar{u}u\rangle_{0}=(-0.272)^3$~GeV$^3$, $\langle \bar{s}s\rangle_{0}=0.8 \langle \bar{u}u\rangle_{0}$~GeV$^3$, $\langle \bar{u}u\rangle_{\rho}=\langle \bar{u}u\rangle_{0}+\frac{\sigma_{\pi N}}{2 m_q}\rho$~GeV$^3$, $\langle \bar{s}s\rangle_{\rho}=\langle \bar{s}s\rangle_{0}+y \frac{\sigma_{\pi N}}{2 m_q}\rho$~GeV$^3$, $\sigma_{\pi N}=0.045$~GeV, $y=0.04\pm0.02;0.066\pm0.011\pm 0.002$, $\langle u^{\dag}g_s\sigma G u\rangle_{\rho}=-0.33 ~GeV^2~\rho$, $\langle s^{\dag}g_s\sigma G s\rangle_{\rho}=-0.33 ~GeV^2~\rho$, $\langle u^{\dag}iD_0u\rangle_{\rho}=0.18~$GeV$~\rho$, $\langle s^{\dag}iD_0 s\rangle_{\rho}=0.18~$GeV$~\rho$, $m_0^2=0.8$~GeV$^2$, $\langle \bar{u}g_s\sigma G u\rangle_{0}=m_0^2 \langle \bar{u}u\rangle_{0}$~GeV$^5$,  $\langle \bar{s}g_s\sigma G s\rangle_{0}=m_0^2 \langle \bar{s}s\rangle_{0}$~GeV$^5$, $\langle \bar{u}g_s\sigma G u\rangle_{\rho}=\langle \bar{u}g_s\sigma G u\rangle_{0}+ 3$~GeV$^2~\rho$~GeV$^5$, $\langle \bar{s}g_s\sigma G s\rangle_{\rho}=\langle \bar{s}g_s\sigma G s\rangle_{0}+ 3$~GeV$^2~y \rho$ ~GeV$^5$, $\langle u^{\dag}iD_0 iD_0 u\rangle_{\rho}=0.03~$GeV$^2~\rho-\frac{1}{8}\langle u^{\dag}g_s\sigma G u\rangle_{\rho}$, $\langle s^{\dag}iD_0 iD_0 s\rangle_{\rho}=0.03~$GeV$^2~y \rho-\frac{1}{8}\langle s^{\dag}g_s\sigma G s\rangle_{\rho}$, $\langle \frac{\alpha_s}{\pi}G^2\rangle_{0}=(0.33\pm0.04)^4~$GeV$^4$ and $\langle \frac{\alpha_s}{\pi}G^2\rangle_{\rho}=\langle \frac{\alpha_s}{\pi}G^2\rangle_{0}-(0.65\pm 0.15)~$GeV$~\rho$ \cite{PhysRevC.45.1881,PhysRevD.98.030001,COHEN1995221,GUBLER20191,Bazavov:2010yq,PhysRevD.88.014513,PhysRevD.90.114504,PhysRevD.93.054502,10.1093/ptep/ptw129,PhysRevC.47.2882,Mishra1993,Sait1998,PhysRevD.87.074503,Belyaev:1982sa,Ioffe:2005ym,Dinter:2011za}. 

The next step is to determine the working regions of the auxiliary parameters $M^2$ and $s_0^*$ that the sum rules depend on. To fix them, the standard criteria of the method such as, mild dependence of  the results on the auxiliary parameters, convergence of the OPE and pole dominance (for details see for instance Refs. \cite{Azizi:2020itk,Azizi:2020yhs}) are used. These requirements lead to the intervals $3 ~GeV^2\leq M^2\leq 5 ~GeV^2$ and $18.3 ~GeV^2\leq s_0^*\leq 20.1~GeV^2$ for the Borel parameter and continuum threshold in the  charmed channel, respectively. 
\begin{figure}[h!]
\label{fig1}
\centering
\begin{tabular}{c}
\epsfig{file=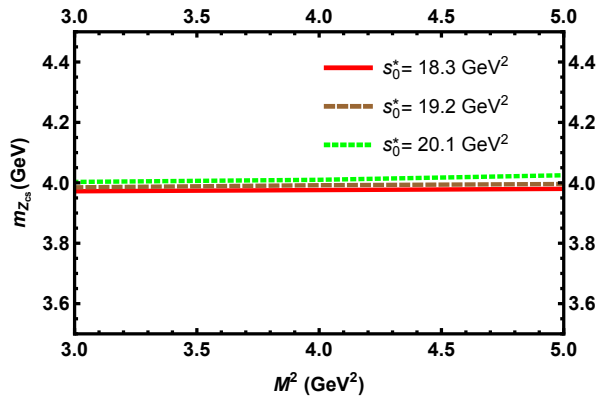,width=0.90\linewidth,clip=}  
\end{tabular}
\caption{The vacuum mass  of $Z_{cs}$ state as function of $M^2$ at different fixed values of continuum threshold.}\label{Fig2}
\end{figure}
\begin{figure}[h!]
\label{fig1}
\centering
\begin{tabular}{c}
\epsfig{file=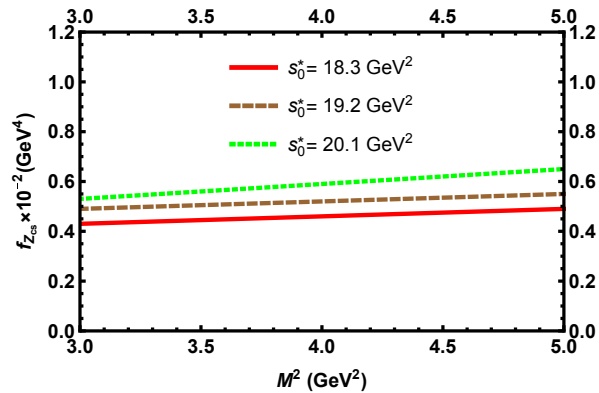,width=0.90\linewidth,clip=}  
\end{tabular}
\caption{The vacuum current coupling  of $Z_{cs}$ state as function of $M^2$ at different fixed values of continuum threshold.}\label{Fig3}
\end{figure}

Making use of all the input parameters and the working windows for the auxiliary parameters, we  plot  the vacuum mass and current coupling of $Z_{cs}$ state (at $\rho \rightarrow 0$ limit)  with respect to $M^2$ at different fixed values of the continuum threshold in figures 1 and 2. From these figures, we see that the vacuum mass depends very weakly on the auxiliary parameters  $M^2$ and $s_0^*$, while the current coupling shows a moderate dependence on them. From the analyses, we  find the values for the vacuum mass and current coupling  as
\begin{eqnarray}
m_{Z_{cs}}&=&3989^{+36}_{-17}~\mbox{MeV},\nonumber  \\
f_{Z_{cs}}&=&(0.54\pm 0.11) \times 10^{-2}~\mbox{GeV$ ^4 $}.
\end{eqnarray}
 Comparing the value of mass obtained in the vacuum with the experimental value of the BESIII Collaboration,  $3982.5^{+1.8}_{-2.6}\pm2.1$ MeV, we see a nice agreement between the two values. The value provided for the current coupling constant in above equation can be served as one of the main inputs in investigation of decay products of the $Z_{cs}$ state.  Note that as it is clear from the above values for the mass and current coupling, the uncertainty in  the value of current coupling is much more compared to that of the mass. This can be attributed to the fact that the mass is found from the ratio of two sum rules in Eq. (\ref{SumR}) that kill the uncertainties of each other and lead to a more accurate result, while the current coupling is obtained using only one sum rule that leads to a relatively large uncertainty. However, all uncertainties remain within the limits allowed by the method.

 At saturation density of the nuclear matter, the mass reduces considerably. At this point, we get 
\begin{eqnarray}
m^*_{Z_{cs}}&=&3846^{+34}_{-16}~\mbox{MeV},\nonumber  \\
f^*_{Z_{cs}}&=&(0.45\pm 0.09 ) \times 10^{-2}~\mbox{GeV$ ^4 $},\nonumber  \\
\Sigma_{\upsilon}&\simeq &1027 ~\mbox{MeV},
\end{eqnarray}
showing that the particle gains a large vector self-energy, referring to the considerable vector repulsion
of this state by the medium at saturated nuclear matter density. These results also indicate that the particle receives a scalar self energy of $ -143 ~\mbox{MeV} $.  This negative shift in the mass shows the scalar attraction of the  $Z_{cs}$ state by the dense medium. 
\begin{figure}[h!]
\centering
\begin{tabular}{c}
\epsfig{file=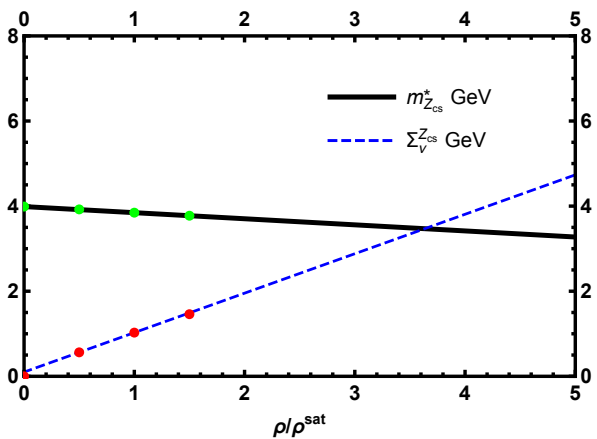,width=0.90\linewidth,clip=}  
\end{tabular}
\caption{The dependence  of $m^*_{Z_{cs}}$ and $\Sigma_{\upsilon}  $ on $\rho/\rho^{sat} $ at mean values of the continuum threshold and Borel
parameter. The green and red points denote the results obtained from the in-medium sum rules, while the solid and dashed lines represent the fit results. }\label{Fig1}
\end{figure}
We show the dependence of   $m^*_{Z_{cs}}$ and $\Sigma_{\upsilon}  $ on $\rho/\rho^{sat} $ at mean values of the continuum threshold and Borel
parameter in figure 3. We should note that our sum rules give reliable results up to roughly $\rho/ \rho^{sat}=1.5 $, after which the results are not decent and  we use some fit functions  to extend the results up to  $\rho=5 \rho^{sat} $, which corresponds to the density of the cores of neutron stars (see \cite{Azizi:2020yhs} for more details).  The fit functions for the in-medium mass, current coupling and vector self-energy are obtained as
 \begin{eqnarray}
\label{fit}
m_{Z_{cs}}^*(x)&=&3989 - 143 x ~\textrm{MeV}, \nonumber \\
f_{Z_{cs}}^*(x)&=&0.54 - 0.09 x ~\textrm{GeV}^4, \nonumber \\
\Sigma_\upsilon^{Z_{cs}} (x) &=&  1027 x ~\textrm{MeV},
\end{eqnarray}
where $x=\rho/\rho^{sat}$. These functions and also the related figure, 3, show that the mass, current coupling and vector self-energy linearly depend on the density: the mass and current coupling  decrease with  increasing in the density, while the vector self-energy  increases with the density.

Finally, we extend our analyses to predict the mass and decay constant of the b-partner of the observed  $ Z_{cs}(3985)^- $ state. Our calculations show that the working windows $9 ~GeV^2\leq M^2\leq 12 ~GeV^2$ and $126 ~GeV^2\leq s_0\leq 130~GeV^2$ satisfy all the requirements of the method. Using the input values in b-channel we get 
\begin{eqnarray}
m_{Z_{bs}}&=&10732^{+97}_{-46}~\mbox{MeV},\nonumber  \\
f_{Z_{bs}}&=&(3.58\pm 0.73) \times 10^{-2}~\mbox{GeV$ ^4 $},
\end{eqnarray}
in vacuum. The future experiments may search for the  charged hidden-bottom open strange state, $ Z^-_{bs} $, with quantum numbers  $ J^{P} = 1^{+}$ and quark composition $ b \bar b s\bar u $ around the $ 10730 ~\mbox{MeV}$ mass. The Belle Collaboration has previously reported the observation  of two narrow structures in the mass spectra of $\mathbf{ \Upsilon}(nS)\pi^+\pi^-, n=1, 2, 3 $ and $ h_b(mP)\pi^+\pi^-, m=1, 2$ pairs called $Z_b(10610)$ and $Z'_b(10650)$ in the decays $\mathbf{ \Upsilon}(5S)\rightarrow \mathbf{ \Upsilon}(nS)\pi^+\pi^-$ and $\mathbf{ \Upsilon}(5S)\rightarrow h_b(mP)\pi^+\pi^-$ \cite{Bondar,Garmash} and assigned the spin-partiy $J^P=1^+$ to these states. The measured masses for these states are
\begin{eqnarray}
m_{Z_b}& = &  (10607.2\pm2.0) ~ MeV, \nonumber \\
m_{Z'_b}& = &  (10652.2\pm1.5) ~ MeV.
\end{eqnarray}
 The Belle Collaboration may search for the strange $ Z_{bs} $ states, as well.

\section{Summary and conclusions}
Inspired by the recent observation of  the charged hidden-charmed open strange $ Z_{cs}(3985)^- $ state by the BESIII Collaboration, we performed a QCD sum rule analysis for the spectroscopic parameters of this state  both in vacuum and a medium with finite density. For its mass in vacuum we obtained $m_{Z_{cs}}=3989^{+36}_{-17}~\mbox{MeV}$, which is in accord with the experimental value, $ m=3982.5^{+1.8}_{-2.6}\pm 2.1~~ \mbox{MeV} $. From this result, we conclude that the assignments $ J^{C} = 1^{+}$ and $ c \bar c s\bar u $  for the quantum numbers and quark structure of this state works well.  We calculated the current coupling constant of this state in vacuum as well, which can be used in investigation of the electromagnetic, weak or strong decays of $Z_{cs}  $ state as an important input parameter. We also estimated the mass and current coupling of the b-partner of this state, $ Z_{bs}$, in vacuum and in the same picture. The obtained value for the mass of this hypothetical   resonance, $ m_{Z_{bs}}=10732^{+97}_{-46}~\mbox{MeV} $, may be checked via different nonperturbative approaches as well as future experiments.

We also reported  the values of mass, current coupling constant and vector self-energy of $Z_{cs}  $ state at the saturation density of a nuclear medium. We investigated the behavior of the spectroscopic parameters of this state in terms of the density and observed that they are linearly changed by increasing in the value of the density: the mass slightly decreases while the vector self-energy considerably increases with respect to the density. Our results may be helpful for future heavy ion collision experiments and those aiming to study the behavior of the exotic states at finite densities.

\begin{acknowledgments}
 The authors thank  TUBITAK for the  partial support provided under the Grant No. 119F094.
   \end{acknowledgments}
   
 \appendix*
\section{The spectral densities corresponding to the structure $g_{\mu\nu}$ used in the calculations}

In this appendix, we collect the explicit forms of the  spectral densities for the $g_{\mu\nu}$ structure. They are obtained as:

\begin{widetext}

\begin{eqnarray}
 \psi_1^{\textrm{pert}}(s,\rho) &= &- \frac{1}{6144\pi^6} \int_0^1 dz \int_0^{1-z} dw \frac{\Big(\varpi s z -\zeta m_c^2(w+z)\Big)^2}{\varrho^4 \varpi \zeta^5}   \Bigg\{ 36 \varrho \zeta^2 m_c(m_s w+m_u z)\Big[-3 \varpi sz \nonumber \\
 &+& \zeta m_c^2 (w+z)\Big] + \zeta^2 m_c^4 z (w+z)^2\Big[z^2+\zeta(6w+z)\Big]+ 5 \varpi^2 s^2 z^3 \Big[z^2+\zeta(14w+z)\Big]  \nonumber \\
 &+& 2 \zeta m_c^2  \Big[-\varpi s z^2 (w+z) [26 \zeta w + 3 \zeta z + 3 z^2] +72 \zeta^2 m_sm_u\Big(z^2+\zeta(w+z)\Big)^2\Big]  \Bigg\}\Theta[L(s,z,w)],
\end{eqnarray}

\begin{eqnarray}
 \psi_1^{\textrm{qq}}(s,\rho) &= & \frac{1}{16\pi^4} \int_0^1 dz \int_0^{1-z}  \frac{dw}{\varrho^3 \zeta^2}   \Bigg\{ \Big[m_c \Big(-s z \varpi + m_c^2 \zeta (w + z)\Big) \Big(-7 s w z \varpi + 
    3 m_c^2 \zeta w (w + z) \nonumber \\
&+& 4 m_c m_u \zeta \varrho\Big)\Big] \langle\bar{s}s\rangle_{\rho}  + \Big[m_c \Big(-s z \varpi + m_c^2 \zeta (w + z)\Big) \Big(-7 s w z \varpi + 
    3 m_c^2 \zeta w (w + z) \nonumber \\
&+& 4 m_c m_s \zeta \varrho\Big)\Big] \langle\bar{u}u\rangle_{\rho} +  4\Big [\varpi m_q m_c p_0 z \Big(2 \varrho \zeta m_c m_u  - 7 \varpi s w z + 5 \zeta m_c^2 w (w+z) \Big) \Big ]  \langle s^{\dagger}s\rangle_{\rho} \nonumber \\
&-&   4\Big [\varpi m_q m_c p_0 z \Big(2 \varrho \zeta m_c m_s  - 7 \varpi s w z + 5 \zeta m_c^2 w (w+z) \Big) \Big ]  \langle u^{\dagger}u\rangle_{\rho}\Bigg\}\Theta[L(s,z,w)],
\end{eqnarray}

\begin{eqnarray}
 \psi_{1}^{\textrm{gg}}(s,\rho) &= & \frac{1}{24\pi^4} \int_0^1 dz \int_0^{1-z}  \frac{m_q  dw}{\varrho^4 \zeta}   \Bigg\{z\Bigg[\varrho \varpi m_c(4p_0^2-3s)(1+12\varpi w)z^2-\varrho\varpi\zeta m_c^2 m_u (w+z) \nonumber \\
 &+&\varrho \zeta m_c^3 (1+12\varpi w)z(w+z) + \varpi^2m_uz\Big[-4 \zeta p_0^2 w +s \big(z^2+\zeta(3w+z)\big)\Big]\Bigg]  \langle s^{\dagger}iD_0s\rangle_{\rho} \nonumber \\
 &+&\Bigg[m_s w\big(-4 \zeta p_0^2 w+sz^2+\zeta s(3w+z)\big)z\varpi^2 + \varrho m_c \Big[\zeta m_c^2(w+z)\big(w^2(1-2z)^2+4w^3z-8z^2\big) \nonumber \\
 &+& m_c m_s \zeta \big(-w^3+w^2(1-6z)-5w(-1+z)z-4(-1+z)z^2\big) \nonumber \\
 &+& z \varpi \Big(3s\big(-w^2(1-z)^2-4w^3z+4z^2\big)+4p_0^2w^2(1+4z\varpi)\Big) \Big]\Bigg]  \langle u^{\dagger}iD_0u\rangle_{\rho} \Bigg\}\Theta[L(s,z,w)],\nonumber \\
 &- & \frac{1}{36864\pi^4} \int_0^1 dz \int_0^{1-z}  \frac{dw}{\varrho^4 \varpi^2\zeta^3}   \Bigg\{ -9\varrho \varpi^2\zeta^2 m_cs z \Big[m_u z\big(4w^2-w(4+z)+4z(1+5z)\big)\nonumber \\
 &+& m_s\big(24w^3+8(-1+z)z^2+3wz(2\zeta+5z)\big) \Big] + 9 \varrho \varpi \zeta^3 m_c^3(w+z)\Big[m_s\big(32w^3-2w(3+w)z\nonumber \\
 &+& (-8+23w)z^2\big)+m_u z\big(12w^2+4z(1+7z)-w(4+9z)\big)\Big]+3\varpi^2sz^2 \Big[ 2 \varrho \zeta^2 m_sm_u\big(w^2\nonumber\\
 &+&2(-1+z)z+w(-1+2z)\big) + \varpi s \big[(-4+35w)z^4+\zeta^2w^2(2w+19z) +\zeta z^2\big(w(-4+22w) \nonumber \\
 &+& 2(-2+18w)z\big)\big]\Big] + \varpi \zeta m_c^2 \Bigg[6 \varrho \zeta^2 m_s m_u \Big[3 \zeta w^2(16w+15z) +z^2\big(91w^2+46(-1+z) \nonumber \\
 &+& w(-43+42z)\big)\Big]-sz\Big[36 \zeta^2 w^3+2 \zeta^2 w^2\Big(48+w\big(-153+w(193+12(-3+w)w\big)\Big)z \nonumber \\
 &+& \zeta z^2 \Big(-24z+w\big(-84+4w(135+w(-272+w(256+3w(-31+9w)))) +(390+w(-1133\nonumber \\
 &+&2w(647+12w(-29+10w))))z+72z^5\big)\Big) +z^4\Big(336w^5+24w^4(-46+13z)+w^2(-1890 \nonumber \\
 &+& (983-360z)z)+w^3(1907+48z(-17+4z))+6(-24+z(21+(-2+z)z))+w(911+z(-602 \nonumber \\
 &+& 193z+12z^3))\Big)\Big] \Bigg] +\zeta^2 m_c^4  \Bigg[ 36 \zeta^4 w^4+3\zeta^2w^3 \Big[44+w(-113+w\big(101+8(-3+w)\big))\Big]z \nonumber \\
 &+& \zeta w z^2 \Big[3w\Big(-60+w(289+w(-488+w(387+4w(-39+11w))))\Big)+\Big(-108+w(831\nonumber \\
 &+& w(-1586+3w(639+4w(-89+29w))))\Big)z\Big] +z^4 \Big[ 576w^624w^5(-85+27z)+w^4(3433\nonumber \\
 &+& 24z(-80+21z))+2(-1+z)(-12+z(27+z(-2+3z)))+w^3(-3379+z(2476\nonumber \\
 &+&24z(-49+11z)))+w^2(1823+z(-1749+z(1025+12z(-36+7z))))+w(-429+z(628\nonumber \\
 &+& z(-401+2z(97+6(-6+z)z))))\Big]\Bigg] \Bigg\}  \Theta[L(s,z,w)] \Big\langle \frac{\alpha_s}{\pi}G^2\Big\rangle_{\rho},
\end{eqnarray}

\begin{eqnarray}
 \psi_{1}^{\textrm{qgq}}(s,\rho) &= & \frac{1}{24\pi^4} \int_0^1 dz \int_0^{1-z}  \frac{dw}{\varrho^4 \zeta^2}   \Bigg\{  z\Bigg[ -m_c^2m_u \zeta^2 \varpi\varrho (w+z) -m_u\zeta  \varpi^2z\Big[4p_0^2\zeta w-s(3\zeta w \nonumber \\
 &+& \zeta z+z^2)\Big] +m_c\varpi z\Big[s\Big(10\zeta^2w^2-\zeta w\big(11+3w(5+12(-2+w)w)\big)z-\zeta\big(-3+w(11\nonumber \\
 &+&36(-3+2w))\big)z^2+\big(3+w(-61-72(-2+w)w)\big)z^3-36\zeta wz^4\Big) +4 p_0^2\big(-12 \zeta^2 w^2\nonumber \\
 &+&3\zeta w(5+w(-5+2w9(1+2w))z+\zeta(-1+w(-19+12w(-3+2w)))z^2+(-1+3w(3\nonumber \\
 &+&8(-2+w)w))z^3+12\zeta wz^4\big)\Big] + m_c^3\zeta\varrho(w+z)\Big[-z+w\Big(-2+15z+2(w+6(-2+w)wz\nonumber \\
 &+& 6\zeta z^2)\Big)\Big]\Bigg]\Bigg\}\Theta[L(s,z,w)] \langle \bar{s}iD_0iD_0s\rangle_{\rho},\nonumber \\
 &+ & \frac{1}{24\pi^4} \int_0^1 dz \int_0^{1-z}  \frac{dw}{\varrho^4 \zeta^2}   \Bigg\{  m_c^3\zeta\varrho(w+z)\big[\zeta w^2+4\zeta^2w^2z+2\zeta(-3+2z^2)\big]\nonumber \\
 &+&m_c\varpi z\Big[(4p_0^2-3s)\zeta^2w^3+(4p_0^2-3s)w^2(1-3w+2w^2)^2z+\zeta w\Big(s(-22+w(7\nonumber \\
 &+& 12(3-2w)w))+4p_0^2(12+w(-7+4w(-3+2w)))\Big)z^2+2\zeta\Big(s(-13+6w(3-2\zeta w))\nonumber \\
 &+&8p_0^2(4+w(-7+2\zeta w))\Big)z^3+4\zeta\Big(s(10-3w^2)+4p_0^2(-8+w^2)\Big)z^4+2(-32p_0^2+7s)z^5\Big] \nonumber \\
 &-&m_c^2m_s\zeta^2\varrho\Big[w^3+5w(-1+z)z+4(-1+z)z^2+w^2(-1+6z)\Big]-m_s \zeta \varpi^2wz\Big[4p_0^2\zeta w\nonumber \\
 &-&s(3\zeta w+\zeta z+z^2)\Big]\Bigg\}\Theta[L(s,z,w)]\langle \bar{u}iD_0iD_0u\rangle_{\rho},\nonumber \\
 &+ & \frac{1}{96\pi^4} \int_0^1 dz \int_0^{1-z}  \frac{dw}{\varrho^4 \zeta^2}   \Bigg\{ m_c^3 \varrho\zeta(w+z)\Big[9\zeta w^2+26\zeta wz+\big(-1+3w(13\nonumber \\
 &+&4(-2+w)w)\big)z^2+12\zeta wz^3\Big] +m_c^2m_u \zeta\Big[6\zeta^3w^3+23\zeta^3w^2z+\zeta^2w(-28+51w)z^2 \nonumber \\
 &+& \zeta^2(-11+68w)z^3+\zeta(-34+63w)z^4+35\zeta z^5+12z^6\Big] + m_u \zeta \varpi^2 z^2 (2p_0^2\zeta w+s\varrho)\nonumber \\
 &-&m_cz\varpi\Bigg[2p_0^2z\Big[-12\zeta^2w^2+3\zeta w\big(5+w(-5+2w)(1+2w)\big)z+\zeta\big(-1+w(-19+12w(-3\nonumber \\
 &+&2w))\big)z^2\big(-1+3w(3+8(-2+w)w)\big)z^3+12\zeta wz^4\Big]+ s \Big[15\zeta^2w^3+77\zeta^2w^2z+\zeta w\Big(-71 \nonumber \\
 &+&4w(40+3(-2+w)w)\big)z^2+\zeta\big(-1+3w(51+4w(-3+2w))\big)z^3+\big(-1+w(95+24(-2\nonumber \\
 &+&w)w)\big)z^4+12\zeta wz^5\Big]\Bigg]\Bigg\}\Theta[L(s,z,w)]\langle \bar{s}g_s\sigma Gs\rangle_{\rho},\nonumber\\
&+ & \frac{1}{96\pi^4} \int_0^1 dz \int_0^{1-z}  \frac{dw}{\varrho^4 \zeta^2}   \Bigg\{  m_c^3 \varrho\zeta(w+z)\Big[\zeta w^2+4\zeta^2w^2z+4\zeta(6+w^2)z^2\nonumber \\
 &+&26z^3\Big] - m_c z\varpi \Bigg[\zeta^2w^3(2p_0^2+s)+w^2(1-3w+2w^2)^2z(2p_0^2+s)\nonumber \\
 &+&\zeta w\Big(2p_0^2(12+w(-7+4w(-3+2w)))+s(-56+w(61+4w(-3+2w)))\Big)z^2\nonumber \\
 &+&2\zeta\Big(4p_0^2(4+w(-7+2\zeta w))+s(-32+w(63+4\zeta w))\Big)z^3\nonumber \\
 &+&2\zeta\Big(4p_0^2(-8+w^2+s(67+2w^2)\Big)z^4+2(-16p_0^2+35s)z^5\Bigg] +m_s\zeta wz\varpi^2(sp_0^2\zeta w\nonumber \\
 &+&s\varrho)-m_c^2\zeta\varrho\Big[6m_u\zeta z\varrho+m_s\Big(w^4-13w(-1+z)z(-1+2z)-2(-1+z)z^2(-7\nonumber \\
 &+&6z-2w^3(1+6z)+w^2(1-25(-1+z)z)\Big)\Big]\Bigg\}\Theta[L(s,z,w)]\langle \bar{u}g_s\sigma Gu\rangle_{\rho},\nonumber \
 \end{eqnarray}
 \begin{eqnarray}
 &+& \frac{1}{96\pi^4} \int_0^1 dz \int_0^{1-z}  \frac{p_0 z dw}{\varrho^4 \zeta^2}   \Bigg\{ -8m_cm_u\zeta^2 \varpi \varrho w +3\varpi^2swz(7\zeta w+\zeta z+z^2)\nonumber \\
 &+& m_c^2 \zeta \Bigg[24w^6z+2(-1+z)^2z^3+3w^5\Big(-5+24(-1+z)z\Big)+w(-1+z)^2z^2\Big(1+24(-1+z)z\Big)\nonumber \\
 &+&w^2(-1+z)z\big(16+3z(15+8z(-5+3z))\big)+w^4\big(30+z(41+24z(-8+5z))\big)\nonumber \\
 &+& w^3\big(-15+z(23+4z(37+6z(-11+5z)))\big)\Bigg] \Bigg\}\Theta[L(s,z,w)]\langle s^{\dag}g_s\sigma Gs\rangle_{\rho},\nonumber \\
 &+ & \frac{1}{96\pi^4} \int_0^1 dz \int_0^{1-z}  \frac{p_0 z dw}{\varrho^4 \zeta^2}   \Bigg\{ -m_c \varrho \varpi \zeta^2 w(2m_s-3m_u)-s\varpi^2z^2(5\zeta w+\zeta z+z^2)\nonumber \\
 &+& m_c^2 \zeta \Bigg[\zeta^2 w^3+\zeta^2w^2(5+4\zeta w)z+\zeta w\Big(-5+2w(9+2w(-5+3w))\Big)z^2 \nonumber \\
 &+& \zeta\Big(-1+4w(3+w(-6+5w))\Big)z^3+2\Big(-1+2w(4+w(-8+5w))\Big)z^4\nonumber \\
 &+&(1+12\zeta w)z^5+4wz^6\Bigg] \Bigg\}\Theta[L(s,z,w)]\langle u^{\dag}g_s\sigma Gu\rangle_{\rho},
\end{eqnarray}

where the following shorthand notations have been used:

\begin{eqnarray}
\varrho&=&\big[w^2 + w (-1 + z) + (-1 + z) z\big], \nonumber \\
\varpi&=&(-1 + w + z),  \nonumber \\
\zeta&=&(-1 + w), \nonumber \\
L[s, w,  z] &=& -\frac{\zeta}{\varrho^2}\Big[-swz\varpi+  mc^2 \Big(w^3 + 2 w (-1 + z) z + (-1 + z) z^2 +  w^2 (-1 + 2z)\Big)\Big].
\end{eqnarray}

\end{widetext}



\end{document}